\documentclass{article}
\usepackage{epsfig}

\date{\today}

\usepackage{amsmath}

\newcommand{\la}{\lambda}

\newcommand{\al}{\alpha}

\newcommand{\ee}{\end{equation}}
\newcommand{\eea}{\end{eqnarray}}
\newcommand{\be}{\begin{equation}}
\newcommand{\bea}{\begin{eqnarray}}

\newcommand{\pa}{\partial}

\newcommand{\vep}{\varepsilon}

\newcommand{\re}[1]{(\ref{#1})}

\begin{document}
\title{New axially symmetric Yang-Mills-Higgs solutions
\\
with negative cosmological constant}

\author{
{\large Eugen Radu}$^{\dagger}$ and
{\large D. H. Tchrakian}$^{\dagger \star}$ \\
$^{\dagger}${\small Department of
Mathematical Physics, National University of Ireland Maynooth,}
\\
$^{\star}${\small School of Theoretical Physics -- DIAS, 10 Burlington
Road, Dublin 4, Ireland }}

\maketitle

\begin{abstract}
We construct numerically new axially symmetric solutions of $SU(2)$
Yang-Mills-Higgs theory in $(3+1)$ anti-de Sitter spacetime.
Two types of finite energy, regular configurations are considered:
multimonopole solutions with magnetic charge $n>1$ and
monopole-antimonopole pairs with zero net magnetic charge.
A somewhat detailed
analysis of the boundary conditions for axially symmetric solutions is presented.
The properties of these solutions are investigated, with a view to
compare with those on a flat spacetime background.
The basic properties of the gravitating generalizations of these
configurations are also discussed.

\end{abstract}
\section{Introduction}
The study of topologically stable
monopole~\cite{'tHooft:1974qc,Polyakov:1974ek,Bogomolny:1975de,
Prasad:1975kr} solutions to the $SU(2)$ Yang-Mills-Higgs (YMH) equations
with adjoint representation Higgs field, in flat spacetime,
is a subject of long standing interest. Two types of solutions to this
system are known: multimonopoles (MM) and monopole--antimonopole (MA)
chains.

The first type comprise topologically stable multimonopole solutions
with topological charge $n$, which for $n\ge 2$ cannot be spherically
symmetric~\cite{Weinberg:1976eq} and possess at most axial symmetry.
For vanishing Higgs self interaction potential, these are solutions of
first order Bogomol'nyi equations and the axially symmetric
multimonopole solutions are
minimal energy and topologically stable. In the presence of a Higgs
potential, i.e. when the Higgs coupling constant $\la >0$, like monopoles
repel~\cite{Manton:1977,KKT} so we would expect that the solutions of
these are saddle points of the energy functional. Ignoringing this
instability, we will include multimonopoles for YMH systems with $\la >0$
in our definition for a MM solution.

Recently, further to these MM solutions, new types of
configurations have been considered, which represent composite states of
monopoles and antimonopoles  in (topologically unstable) equilibrium.
The existence of such solutions was first proven by Taubes for a model
featuring no Higgs potential \cite{Taubes:1982ie}. Notwithstanding the
absence of the Higgs potential, such solutions are not self--dual and
their energies exceed the Bogomol'nyi bound.
Kleihaus and Kunz~\cite{Kleihaus:2000sx} (see also \cite{map})
constructed numerically an axially symmetric solution with zero net
magnetic charge, consisting of a  MA  pair for the
YMH system with and without a Higgs potential.
More complicated configurations consisting of MA chains and vortex
rings were constructed subsequently~\cite{Kleihaus:2004is, Shnir:2004yh}.

The main difference between the MM and MA solutions in flat space is
characterised by their distinct boundary conditions at infinity. They have
different boundary conditions also at the origin.

When gravity is switched on, or equivalently when the YMH system is
considered on a curved background, the arguments leading to the saturation
of a topological lower bound and hence to the first order Bogomol'nyi
equations disappear and as a result only solutions to the second order
Euler--Lagrange equations can be sought.

Naturally, the boundary conditions employed play a crucial role in the
properties of these solutions. Even in flat spacetime, the main difference
between the topologically stable axially symmetric MM solutions on the one
hand, and the MA type on the other, is a result of the
imposition of different boundary conditions. Otherwise, both types of
solutions obey the same axially symmetric YMH equations~\footnote{In the
absence of a Higgs potential the MM solutions obey the first order
Bogomol'nyi equations, while the MA solutions obey the second order
Euler--Lagrange equations.}. 
However, to date, all {\it regular} axially symmetric $SU(2)$ YMH
configurations of the types mentioned above have been studied on
a(n asymptotically-) flat spacetime.
It is therefore important to analyse the effects on these solutions,
of different asymptotic structures of the spacetime.

To this end, we are motivated to study the properties of these
configurations for spacetimes with a cosmological constant $\Lambda$,
in particular.  In what follows we consider the case of a negative
$\Lambda$, correspoding to an anti-de Sitter (AdS) geometry.
This maximally symmetric
spacetime has recently enjoyed a considerable amount of interest,
motivated mainly by the proposed correspondence between physical effects
associated with gravitating fields propagating in AdS spacetime and those
of a conformal field theory (CFT) on the boundary of AdS spacetime
\cite{Witten:1998qj,Maldacena:1997re}. This adds further justification
to the study of classical solutions of various field theories
in AdS, the YMH case at hand being an obvious example.

Prominent amongst the remarkable features displayed by
solutions of the gravitating Yang-Mills (YM)
system with $\Lambda<0$ is the existence
of {\bf stable} particle-like and black hole
solutions~\cite{Winstanley:1998sn, Bjoraker:2000qd}. This results from
a particular behaviour of the YM field asymptotically, which feature
is present also for the case of a {\bf fixed} AdS background~
\cite{Boutaleb-Joutei:1979va, Hosotani:2001iz}
\footnote{In Ref.~\cite{Boutaleb-Joutei:1979va}, a solution to the
$SU(2)$ YM system in a
fixed dS background is found in closed form. Passing from dS to AdS
by an appropriate change of sign in this solution, one discovers the same
interesting feature mentioned above, which was discovered in
\cite{Winstanley:1998sn, Bjoraker:2000qd} for the gravitating case.}.
One aim of this work is to inquire to what extent
the features of YM systems with $\Lambda<0$, persist also for YMH
systems?

Here we present numerical arguments for the existence of
both MM  and MA configurations, both in a fixed AdS spacetime
background, and in the fully gravitating case. While the static solutions
of the YM system are drastically different from those of the EYM ones, in
the case of the YMH system the solutions with $\Lambda < 0$ (with or
without gravity) are found to be very similar to those of the
corresponding Minkowski spacetime configurations. (The situation is the
same also for asymptotically flat EYMH.)

As found in \cite{gam, Kleihaus:2000hx} for $\Lambda=0$, when gravity is
coupled to YMH theory, a branch of gravitating solutions emerge from the
flat spacetime configurations. This branch extends up to some maximal
value of the gravitational strength.
The YMH system with a negative cosmological constant has recently been
studied~\cite{Lugo:1999fm,Lugo:1999ai}, for the spherically
symmetric, unit magnetic charge gravitating monopole. It was
found~\cite{Lugo:1999fm} that unlike in flat spacetime, in a fixed AdS
background there are no analytic solutions that might be used as a guide.
Moreover, when a cosmological constant is included (no matter how small
this constant is) no solution close to the flat space 
BPS configuration can be found.
Numerically constructed monopole solutions have been exhibited in
\cite{Lugo:1999ai}, where the effects of gravity are also included.
A distinctive feature of AdS solutions concerns the asymptotic behaviour
of the fields.  When $\Lambda<0$, the Higgs field approaches its vacuum
expectation value faster than in the flat space case. The radius of the
monopole core decreases, 
as the magnetic field concentrates near the origin.

We aim to extend this analysis for axially symmetric solutions,
by studying the basic properties for both types of
configurations, MM solutions with nonvanishing magnetic charge and
MA pairs with a zero net magnetic charge.
As expected, we find that the basic properties of
the  gravitating AdS configurations configurations
are similar to the  asymptotically flat spacetime counterparts.

The paper is structured as follows: in the next Section we present the
model and subject the system to axial symmetry. The imposition of axial
symmetry is carried out in a uniform manner to include the two distinct
boundary conditions pertaining to multimonopole and monopole--antimonopole
solutions. Numerical solutions in a fixed AdS background are constructed
in Section 3 for the two different sets of boundary conditions.
Section 4 contains a  discussion of the main properties of the gravitating
counterparts of these solutions.
We conclude with Section 5 where the results are summarised.

\section{The model}
\subsection{Action principle and field equations}
We consider the action for an $SO(3)$  YM  field $A_{\mu}^a$
coupled to a triplet Higgs field $\Phi^a$ with the usual potential 
$V(\Phi)=\frac{\lambda}{4} (|\Phi^a|^2 - \eta^2)^2$
\begin{eqnarray}
\label{lag0}
S\stackrel{\rm def}=\int\,d^4x\,{\cal L}=
\int d^{4}x\sqrt{-\mbox{det}\,g}\left[
\frac{1}{4g^2}F_{\mu \nu}^aF^{\mu \nu}_a+
\frac{1}{2}D_{\mu}\Phi^aD^{\mu}\Phi_a+
V(\Phi)\right],
\end{eqnarray}
where the field strength tensor and the covariant derivative are
defined as
\[
F_{\mu \nu}^a=\partial_{\mu}A_{\nu}^a-\partial_{\nu}A_{\mu}^a+
\vep^{abc}A_{\mu}^b\,A_{\nu}^c\quad,\quad
D_{\mu}\Phi^a=\partial_{\mu}\Phi^a+\vep^{abc}A_{\mu}\Phi^c\,,
\]
and $g={\rm det}\,g_{\mu\nu}$, $g_{\mu\nu}$ being the metric of a fixed
anti de Sitter (AdS) background. Subjecting the action \re{lag0} to
the variational principle results in the Euler--Lagrange
equations, and varying the Lagrangian ${\cal L}$ with respect
to the metric $g^{\mu \nu}$ yields the energy-momentum tensor
\begin{eqnarray}
\label{tensor}
T_{\mu\nu} =
F_{\mu \alpha}^a F_{\nu \beta}^a g^{\alpha \beta}
-\frac{1}{4}g_{\mu \nu}F^a_{\alpha \beta}F^{a}_{\gamma \delta }g^{\alpha\gamma}g^{\beta\delta}
+  \frac{1}{2}D_{\mu}\Phi^a D_{\nu}\Phi^a
-\frac{1}{4}g_{\mu \nu}(D_{\alpha}\Phi^a D^{\alpha}\Phi^a +
V(\Phi)g_{\mu\nu}).
\end{eqnarray}

We are interested in static, purely magnetic ($A_t=0$),
axially symmetric finite energy solutions of the Euler--Lagrange
equations. The energy density of a solution of the YMH equations
is given by the $tt-$component of the energy-momentum tensor;
integration over all space yields the total mass/energy
\begin{eqnarray}
\label{total-energy}
M = -\int T_{t}^{t}\sqrt{-g} d^{3}x.
\end{eqnarray}
For the mass integral to converge, each term in the integrand of
\re{total-energy} must vanish at large $r$. This will give the required
asymptotic behaviour of the gauge and Higgs field.
The additional requirements to have a finite, locally integrable energy
density  impose boundary conditions at the origin and on the $z-$axis.

The static energy functional arising from \re{total-energy} is bounded
from below, in flat spacetime, by the topological charge
\be
\label{topch}
Q_m=\frac{1}{4\pi}\int\,d^3x\,\vep_{ijk}\,F_{ij}^a\,D_k\Phi^a=
\frac{1}{4\pi}\int\,dS^k\,\vep_{ijk}\,F_{ij}^a\,\Phi^a\,,
\ee
which is the magnetic monopole charge, whose definition is valid also
in curved spacetime.

\subsection{Imposition of axial symmetry}

We start by stating the line element for the
fixed AdS background
\begin{eqnarray}
\label{AdS}
ds^2= g_{\mu \nu} dx^\mu dx^\nu=\frac{d r^2}
{1-\frac{\Lambda}{3}r^2}+
r^2 (d \theta^2+ r^2 \sin ^2 \theta d\varphi^2)
-\left(1-\frac{\Lambda}{3}r^2\right) dt^2,
\end{eqnarray}
where $r,~\theta,~\varphi$ are the usual spherical coordinates.
The  YMH  system is defined by \re{lag0},
with a metric tensor $g_{\mu \nu}$ given by \re{AdS}.

We proceed in this section to present the
usual axially symmetric Ansatz of Rebbi
and Rossi~\cite{Rebbi:1980yi}, adapted in particular to solutions
with boundary conditions supporting monopole-antimonopoles. Particular
attention is given to finding the relevant asymptotic values.
Note also  that this analysis is valid
in the limit $\Lambda=0$.

Expressing the $SU(2)$ gauge connection $A_{\mu}$ and the algebra
valued Higgs field $\Phi$ in terms of components labeled by the algebra
indices
\[
A_{\mu}=A_{\mu}^a\frac{\tau_a}{2}\quad ,\quad
\Phi=\Phi^a\frac{\tau_a}{2}\quad ,
\quad a=\alpha,3\equiv\alpha,z\quad ;\quad \alpha=1,2\equiv x,y\ ,
\]
and splitting up the spacelike coordinates
$x_{\mu}=(x_i,x_3)\equiv(x_i,x_3)$, with $x_i=(x_1,x_2)\equiv(x,y)$,
(where $r=\sqrt{\rho^2+z^2},\ \theta=\arctan{\frac{\rho}{z}})$
the axially symmetric Rebbi--Rossi Ansatz can be expressed as follows
\bea
A_i^{\alpha}&=&-a_{\rho}\,\hat x_i\,(\vep n)^{\alpha}+
\left(\frac{\chi^1}{\rho}\right)\,(\vep\hat x)_i\,n^{\al}\nonumber\\
A_i^3&=&\left(\frac{n+\chi^2}{\rho}\right)\,(\vep\hat x)_i\label{Gauge}\\
A_3^{\al}&=&-a_z\,(\vep n)^{\alpha}\quad ,\quad A_3^3=0\nonumber
\eea
for the gauge connection, and
\be
\label{higgs}
\Phi^{\al}=\eta\,\phi^1\,n^{\al}\quad ,\quad\Phi^3=\eta\,\phi^2
\ee
for the Higgs field, $\eta$ being the VEV of the Higgs field, with inverse
dimension of a length.
The six functions in \re{Gauge}-\re{higgs} depend
on $\rho=\sqrt{|x_{\al}|^2}$ and $x_3=z$, the unit vector in the azimuthal
plane is
\[
n^{\al}=(\cos n\phi ,\sin n\phi)\,,
\]
and $\vep_{AB}$ in \re{gauge} is the antisymmetric Levi-Civita symbol.
This ansatz contains also an integer $n$, which the winding number.

Subject to the Ansatz \re{Gauge} and \re{higgs}, the static energy
density functional \re{lag0} reduces to a 2 dimensional system in
$(\rho,z)$ space which exhibits a residual $U(1)$ gauge invariance.
The two components of the gauge connection are $(a_{\rho},a_z)$
appearing in the Ansatz \re{Gauge}, and the gauge arbitrariness can be
removed by the gauge condition
\be
\label{gaugecond}
\pa_{\rho}\,a_{\rho}+\pa_z\,a_z=0\,.
\ee
Since it is more convenient to work in spherical coordinates, we replace the
two functions $(a_{\rho},a_z)$ by $(a_r,a_{\theta})$ according to
\[
a_{\rho}=a_r\sin\theta+a_{\theta}\frac{\cos\theta}{r}\quad ,\quad
a_z=a_r\cos\theta-a_{\theta}\frac{\sin\theta}{r}\ .
\]
With this replacement, and denoting $\chi^A=(\chi^1,\chi^2)$, one can write
\bea
\label{redlag}
{\cal L}&=&\sqrt{-g}
\Bigg\{ g^{rr}g^{\theta \theta} f_{r\theta}^2
+g^{\varphi \varphi}\left( g^{rr}|{\cal D}_r\chi^A|^2
+g^{\theta \theta}|{\cal D}_{\theta}\chi^A|^2\right)
\nonumber\\
&+&\eta^2\left[  g^{rr}|{\cal D}_r\phi^A|^2
+g^{\theta \theta} |{\cal D}_{\theta}\phi^A|^2
+g^{\varphi \varphi}\left(\vep_{AB}\phi^A\chi^B\right)^2\right]
+V(\phi^A)
\Bigg\},
\eea
which is manifestly a $U(1)$ gauged Higgs model, with $U(1)$ connection
$(a_r,a_{\theta})$ and Higgs field doublets $\chi^A=(\chi^1,\chi^2)$,
$\phi^A=(\phi^1,\phi^2)$, and with the correspoding $U(1)$ curvature and
covariant derivatives in \re{redlag} defined by
\bea
f_{\mu\nu}&=&\pa_{\mu}a_{\nu}-\pa_{\nu}a_{\mu}\nonumber\\
{\cal D}_{\mu}\chi^A&=&\pa_{\mu}\chi^A+a_{\mu}\vep^{AB}\chi^B\nonumber\\
{\cal D}_{\mu}\phi^A&=&\pa_{\mu}\phi^A+a_{\mu}\vep^{AB}\phi^B\nonumber
\eea
having denoted $\mu=r,\theta$, not to be confused with the label $\mu$
in \re{lag0}.

We now consider the asymptotic conditions in the $r\gg 1$ region that
must be satisfied for finite energy, and the conditions of analyticity
at $r=0$ and on the $z$-axis.

It is immediately obvious from the Ansatz \re{Gauge}, in this gauge, that
when $\rho\to 0$
\be
\label{chi-rho0}
\lim_{\rho\to 0}\chi^1=0\quad ,\quad \lim_{\rho\to 0}\chi^2=-n\,,
\ee
which will be used presently to find the asymptotic values of the fields
in the $r\gg 1$ region. In the same ($\rho\to 0$) limit, differentiability
sets the following condition on the Higgs field function $\phi^1$ in
\re{higgs}
\be
\label{phi-rho0}
\lim_{\rho\to 0}\phi^1\propto\rho^n\quad,\quad{\rm i.e.}\quad
\phi^1|_{r=0}=0\quad{\rm and}\quad\phi^1|_{\theta=0,\pi}=0\,.
\ee
but no condition on the other Higgs field function $\phi^2$. Thus the
zeros of the Higgs field, i.e. when both $|\phi^1|^2=0$ and $|\phi^2|^2=0$
will not necessarily occur at the origin in general.
\subsection{Asymptotics}
In addition to the particular conditions
\re{chi-rho0} and \re{phi-rho0} to be satisfied on the $z$-axis, the
comprehensive set of asymptotic values will now be stated, both in the
$r\gg 1$ and $r\ll 1$ regions, in that order
(here also the analysis remains valid
in the flat space limit).

\medskip
\noindent
\underline{Asymptotics in the $r\gg 1$ region}

\smallskip
In the $r\gg 1$ asymptotic region on the other hand, the effect of
a symmetry breaking Higgs potential, whether explicitly included in the
action or not, results in the following finite energy condition on the
Higgs field \re{higgs}
\be
\label{phi-rinfty}
\lim_{r\to\infty}\phi^1=\sin m\theta\quad,\quad
\lim_{r\to\infty}\phi^2=\cos m\theta\,,
\ee
$m$ being an integer. For $m=1$, \re{phi-rinfty} are just the usual
boundary conditions applying to (multi-) monopole solutions, and with $n=1$ in
\re{higgs}, they are the boundary conditions for the spherically symmetric
unit monopole. For $m\ge 2$, the solutions describe
monopole--antimonopole chains, with magnetic charges equal to the
winding of the asymptotic Higgs field \re{phi-rinfty},
\be
\label{mapch}
Q_m=2\pi n\,[1-(-1)^{m}]\ .
\ee

Finiteness of the energy requires that both terms $|{\cal D}_r\phi^A|^2$
and $|{\cal D}_{\theta}\phi^A|^2$ vanish at infinity, yielding
\be
\label{asyma}
\lim_{r\to\infty}a_r=0\quad,\quad\lim_{r\to\infty}a_{\theta}=-m\,,
\ee
while that for the last term in \re{redlag} describing the interaction
between $\chi^A$ and $\phi^A$ yields
\be
\label{asymint}
(\vep\phi)_A\,\chi^A=0
\ee
in this limit, where we have expressed \re{phi-rinfty} as
\be
\label{t}
\lim_{r\to\infty}\phi^A=t^A\,,
\ee
in terms of the unit 2-vector $t^A=(\sin m\theta,\cos m\theta)$. The
$\theta$ derivative of \re{t} gives the condition
\be
\label{tcond}
(\vep t)_A\,\pa_{\theta}\chi^A=m\,t_A\,\chi^A\,,
\ee
which we will find convenient to use presently.

The corresponding condition for the vanishing of $|{\cal D}_r\chi^A|^2$
in \re{redlag} at infinity implies that $\chi^A\to\rm{const.}$ in this
limit, but the actual asymptotic values of $\chi^A$ cannot be determined
without analysing the corresponding Euler--Lagrange equations, since
finiteness of energy does not require that the term
$|{\cal D}_{\theta}\chi^A|^2$ in \re{redlag} vanish at infinity. To
leading order  in this limit, this equation is
\[
\sin\theta\,{\cal D}_{\theta}\left(
\frac{1}{\sin\theta}{\cal D}_{\theta}\chi\right)^A=0\,.
\]
Contracting this equation with the unit vector $t^A$ and using
\re{asymint} and \re{tcond}, it reduces to
\be
\label{1storder}
\frac{d}{d\theta}\,\ln\frac{t_A\pa_{\theta}\chi^A}{\sin\theta}=0\,,
\ee
which is immediately integrated to give
\be
\label{soln}
t_A\,\pa_{\theta}\chi^A=c_1\sin\theta\,,
\ee
$c_1$ being an integration constant. \re{soln} and \re{asymint} together
now result in
\[
\frac{d}{d\theta}\,(t_A\,\chi^A)+c_1\sin\theta=0\,,
\]
yeilding finally
\be
\label{chiasym}
\chi^A=(c_2+c_1\cos\theta)\,t^A\,,
\ee
$c_2$ being the second integration constant.

It is easy to check that the asymptotic fields \re{chiasym} as they
stand, result in the cancellation of the singularities due to
$\frac{1}{\sin^2\theta}$ in \re{redlag}. To fix the integration constants
$c_1$ and $c_2$ in \re{chiasym}, additional constraints are needed. To
this end we recall the
analyticity conditions \re{chi-rho0} on the $z$-axis. Substituting
\re{chiasym} in \re{chi-rho0}, we find that the first member of the
latter is identically satisfied while the second yieds
\bea
\chi^2|_{\theta=0}&=&(c_2+c_1)=-n\nonumber\\
\chi^2|_{\theta=\pi}&=&(-1)^m\,(c_2-c_1)=-n\,,\nonumber
\eea
yielding the values of $c_1$ and $c_2$ to be used in \re{chiasym},
\bea
c_1=0\quad,\quad c_2=-n\,,\quad &m&\ \rm{even}\label{c12odd}\\
c_1=-n\quad,\quad c_2=0\,,\quad &m&\ \rm{odd}\,.\label{c12even}
\eea
\medskip
\noindent
\underline{Asymptotics in the $r\ll 1$ region}

\smallskip
The $r\ll 1$ region is a special case of the $\rho=0$ line, namely the
$z$-axis. Starting from the constraints of analyticity on the $z$-axis,
we have already \re{chi-rho0}-\re{phi-rho0}. We restate these asymptotic
behavioures in the $\rho\to 0$ limit, along with the other
such conditions implied by the Ansatz \re{Gauge}, although we will be
concerned almost exclusively with the actual asymptotic values. For the
Higgs field funtions $\phi^A$, $A=1,2$, these are
\be
\label{rho0higgs}
\phi^1=a_1(z)\rho^n\quad,\quad\phi^2=a_2(z)\,,
\ee
and for the gauge field functions $(a_{\rho},a_z)$ and $\chi^A$
$A=1,2$,
\be
\label{rho0gauge}
a_{\rho}=b_1(z)\rho^{n+1}\quad,\quad a_z=b_2(z)\rho^n\quad,\quad
\chi^1=c_1(z)\rho^{n+2}\quad,\quad\chi^2=-n+c_2(z)\rho^2\,.
\ee
It is clear from \re{rho0gauge} that the asymptotic values of these
functions are simply stated as
\be
\label{rho0gaugevalue}
a_{\rho}|_{r=0}=0\quad,\quad a_z|_{r=0}=0\quad,\quad
\chi^1|_{r=0}=0\quad,\quad\chi^2|_{r=0}=-n\,.
\ee
To state the asymptotic values of the functions $(\phi^1,\phi^2)$ on the
other hand is much harder. It is well known that for axially symmetric
multimonopoles \cite{fhp}, with $m=1$, these are
\be
\label{rho0higgsMM}
\phi^1|_{r=0}=0\quad,\quad\phi^2|_{r=0}=0\,,
\ee
while for monopole-antimonopole-chain solutions with $m\ge 2$, the
situation is more complex. Clearly, for monopole-antimonopole-chains on
the $z$-axis, the function $a_2(z)$ in \re{rho0higgs} must have zeros
away from $r=0$, precluding the second member of \re{rho0higgsMM}. To
date, where these zeros occur has not been determined analytically,
but are found through the numerical
process~\cite{Kleihaus:2004is,Kleihaus:2000hx}.

For monopole-antimonopoles \re{rho0higgsMM} is inadequate, so from
\re{rho0higgs} we infer the general, weaker, boundary conditions
\be
\label{rho0higgsMA}
\phi^1|_{r=0}=0\quad,\quad\pa_{\rho}\phi^2|_{\rho=0}=0\,.
\ee
We note that both \re{rho0higgsMM} and \re{rho0higgsMA} are consistent
with the conclusions of \cite{Houston:aj}, namely that theere is only
one zero of the Higgs field for an {\it axially symmetric} MM solution,
and that if there are distinct zeros on the $z$-axis, either {\it axially
symmetry} is violated or the charges corresponding to the distinct
zeros have differening signs.

\medskip
\noindent
\underline{The parametrisation of Kleihaus and Kunz}

\smallskip
This particular parametrisation of the Rebbi and Rossi Ansatz used by
these authors, see e.g. \cite{Hartmann:2001ic}, is very convenient and
is employed in most works on
axially symmetric YMH systems.  Since the numerical integrations will be
carried out for the variables $(r,\theta)$ and not $(\rho,z)$, it is
useful to employ a parametrisation
for which the Ansatz \re{Gauge}-\re{higgs} agrees with the unit
magnetic charge monopole for $n=1$. The functions $(H_1,H_2,H_3,H_4)$
parametrising the gauge field are related to the functions in the axially
symmetric Ansatz \re{Gauge} as follows
\begin{eqnarray}
\label{ans11}
H_1&=&r(a_{\rho}\sin\theta+a_z\cos\theta)\nonumber\\
1-H_2&=&r(a_{\rho}\cos\theta-a_z\sin\theta)\nonumber\\
&&\label{KKG}\\
n\sin\theta  H_3&=&\chi^1\sin\theta+(n+\chi^2)\cos\theta\nonumber\\
n\sin\theta (1-H_4)&=&\chi^1\cos\theta-(n+\chi^2)\sin\theta\,,
\nonumber
\end{eqnarray}
and the functions $(\Phi_1,\Phi_2)$ parametrising the Higgs field Ansatz
\re{higgs},
\begin{eqnarray}
\label{ans12}
\label{KKH}
\Phi_1=\phi^1\sin\theta+\phi^2\cos\theta\quad,\quad
\Phi_2=\phi^1\cos\theta-\phi^2\sin\theta\,.
\end{eqnarray}

In terms of this parametrisation, the boundary conditions for the
monopole--antimonopole solutions at $r\to\infty$, can be read off
\re{asyma}, \re{chiasym}-\re{c12even}, and \re{KKG}. For odd $m$ these are
\begin{eqnarray}
\label{oddm}
H_1=0,~~H_2=-(m-1),~~
H_3=\frac{\cos \theta }{ \sin \theta}[\cos(m-1)\theta -1],~~
H_4=-\frac{\cos \theta}{\sin \theta}\sin(m-1)\theta\,.
\end{eqnarray}
and for even $m$
\begin{eqnarray}
\label{evenm}
H_1=0,~~H_2=-(m-1),~~
H_3=\frac{1}{\sin \theta}[\cos(m-1) \theta  -\cos \theta ],~~
H_4=-\frac{  \sin(m-1)\theta }{\sin \theta},
\end{eqnarray}
while
\begin{eqnarray}
\label{BC-H}
\Phi_1=\cos (m-1) \theta,~~\Phi_2=\sin (m-1) \theta,
\end{eqnarray}
for any value of $m$.

The corresponding boundary conditions for multimonopoles are simply
given by the set for odd $m$, \re{oddm}, with $m=1$.

The boundary conditions for the gauge field
functions at $r\to 0$ can be read off \re{rho0gaugevalue},
for both multimonopoles   and monopole-antimonopoles. The Higgs
field boundary conditions for MM's and MA's
on the other hand, are different, being stated respectively by
\re{rho0higgsMM} and \re{rho0higgsMA}. Using the relations
\re{KKG}-\re{KKH}, \re{rho0gaugevalue} and \re{rho0higgsMM} yield the
required boundary values at $r=0$ for multimonopoles
\begin{eqnarray} 
\label{r0MM}
H_{1}=H_3=0, ~~
H_{2}=H_{4}=1, ~~
\Phi_{1}=\Phi_{2}=0.
\end{eqnarray}
For monopole-antimonopoles, the less stringent condition \re{rho0higgsMA}
must be used instead of \re{rho0higgsMM}. Following
\cite{Kleihaus:2000hx,Kleihaus:2004is}
we state the boundary values for monopole-antimonopoles as
\begin{eqnarray}
\label{r0MA}
H_{1}=H_3=0, ~~
H_{2}=H_{4}=1,~~
\cos \theta\, \pa_r\Phi_{1}-\sin \theta\,\pa_r \Phi_{2}=0,~~
\sin \theta\, \Phi_{1}+\cos \theta\, \Phi_{2}=0\,,
\end{eqnarray}
the Higgs part of which is a weaker version of \re{rho0higgsMA}.

The above parametrisation is in fact employed in the literature for
the numerical construction of axially symmetric MM solutions, e.g. in
\cite{Hartmann:2001ic} and also for gravitating YM solutions
\cite{Kleihaus:1997mn}.
In the corresponding construction of
axially symmetric MA solutions however, these authors use a different
parametrisation in Ref.~\cite{Kleihaus:2000sx}, as well as in
\cite{Kleihaus:2000hx}, where they have constructed the gravitating
counterparts of the latter. However, the parametrisation used in the MA
works \cite{Kleihaus:2000sx,Kleihaus:2000hx} is equivalent to that of
\cite{Hartmann:2001ic}, which is what we have finally used in the present
work, unifying the treatments of axially symmetric MM and MA solutions.

\newpage
\setlength{\unitlength}{1cm}
\begin{picture}(18,7.5)
\centering
\put(2,0){\epsfig{file=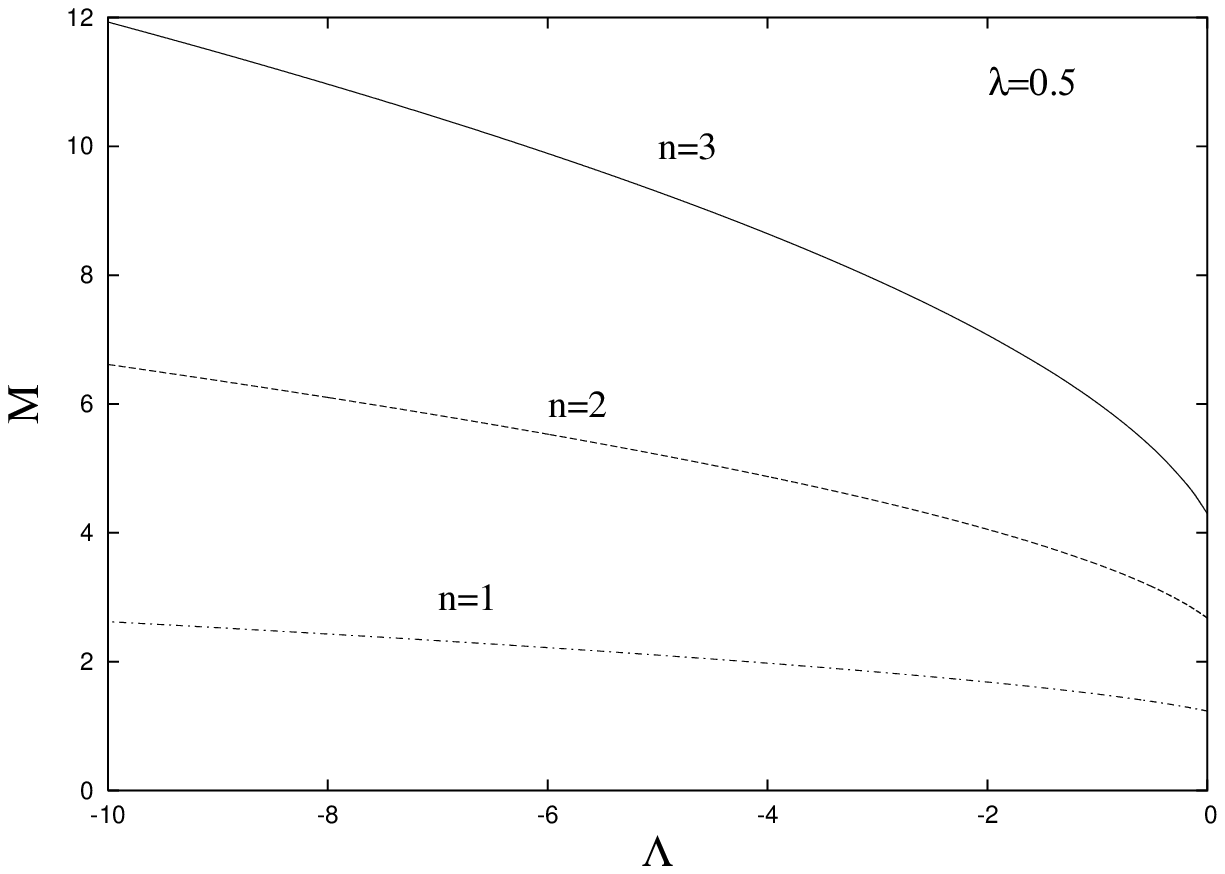,width=12cm}}
\end{picture}
\\
\\
{\small {\bf Figure 1.}
The mass of solutions is plotted
as a function of the cosmological constant for three values of the
winding number $n$ and Higgs self-coupling constant $\lambda=0.5$.
Here and in Figures 2  and 4a the mass is given in units $4\pi \eta/g$.}
\section{Solutions in a fixed AdS background}
We have performed our numerical computations using the parametrisation of
Kleihaus and Kunz described above. All numerical calculations were
performed with the software package CADSOL/FIDISOL, based on the
Newton-Raphson method \cite{FIDISOL}.

\subsection{Multimonopole solutions}
We start by discussing the simplest example of axially symmetric solution,
describing multimonopoles  containing only magnetic charges with the
same sign ($m=1$). The suitable boundary conditions for the Higgs field at
infinity are $\Phi_1=1,~\Phi_{2}=0$,
while, in the same limit the YM potentials should vanish $H_i=0$.

Given the parity reflection symmetry satisified by the ansatz
(\ref{ans11})-(\ref{ans12}), we need to consider solutions
only in the region $0 \leq \theta \leq \pi/2$;
on the $z$- and $\rho$-axis the functions $H_1, H_3, \Phi_2$ and the
derivatives $\partial_\theta H_2,\partial_\theta H_4 $ and
$\partial_\theta \Phi_1$ are to vanish. To fix the residual abelian gauge
invariance we choose the usual gauge condition
(\ref{gaugecond}) which reads in terms of $H_i$ functions
\begin{eqnarray}
\label{gauge}
r \partial_r H_1 - \partial_\theta H_2 = 0.
\end{eqnarray}
For these boundary conditions,
the configurations with $n=1$ corresponds
to spherically symmetric monopoles with 
$H_1=H_3=\Phi_2=0,~H_2=H_4=w(r),
~\Phi_1=H(r),$
and generalize the 't Hooft-Polyakov solution
for a negative cosmological constant.
Axially symmetric solutions are found for $n=2,3,\dots$.

Dimensionless coordinates and Higgs field are obtained by rescaling
\begin{eqnarray}
\label{resc}
r \to r /\eta,~~~\Lambda \to \Lambda \eta^2,~~\Phi \to \eta \Phi.
\end{eqnarray}

Within this ansatz and gauge choice, we have solved the resulting set of
six non-linear coupled partial differential equations,
finding solutions for any considered value of the cosmological constant.

\newpage
\setlength{\unitlength}{1cm}
\begin{picture}(18,7.5)
\centering
\put(2,0){\epsfig{file=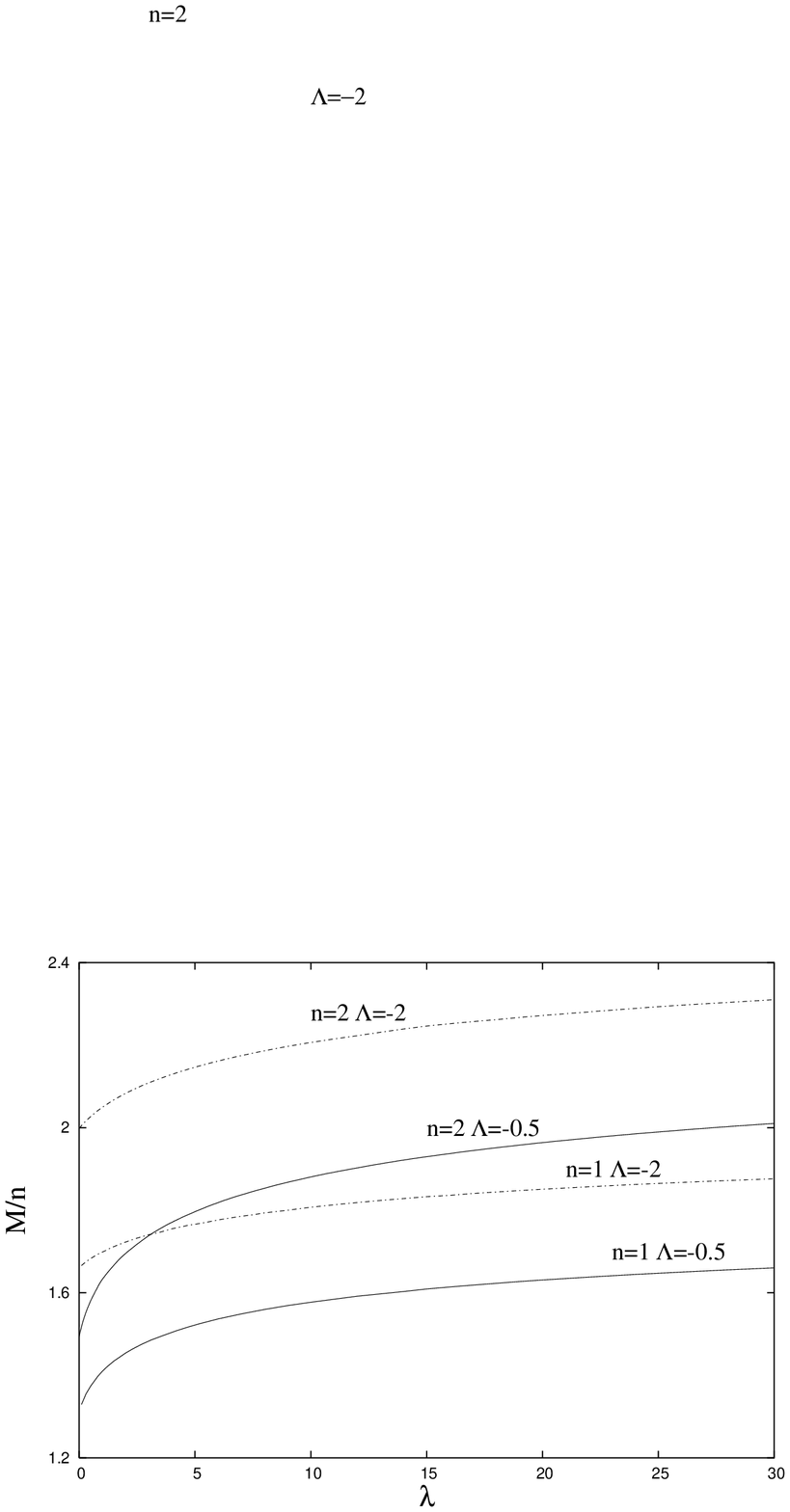,width=12cm}}
\end{picture}
\\
\\
{\small {\bf Figure 2.}
The mass per topological charge  is plotted
as a function of Higgs self-coupling constant $\lambda$ for MM solutions with $\Lambda=-0.5,~-2$
and $n=1,~2$.}
\\
\setlength{\unitlength}{1cm}
\begin{picture}(18,7.5)
\centering
\put(3,0){\epsfig{file=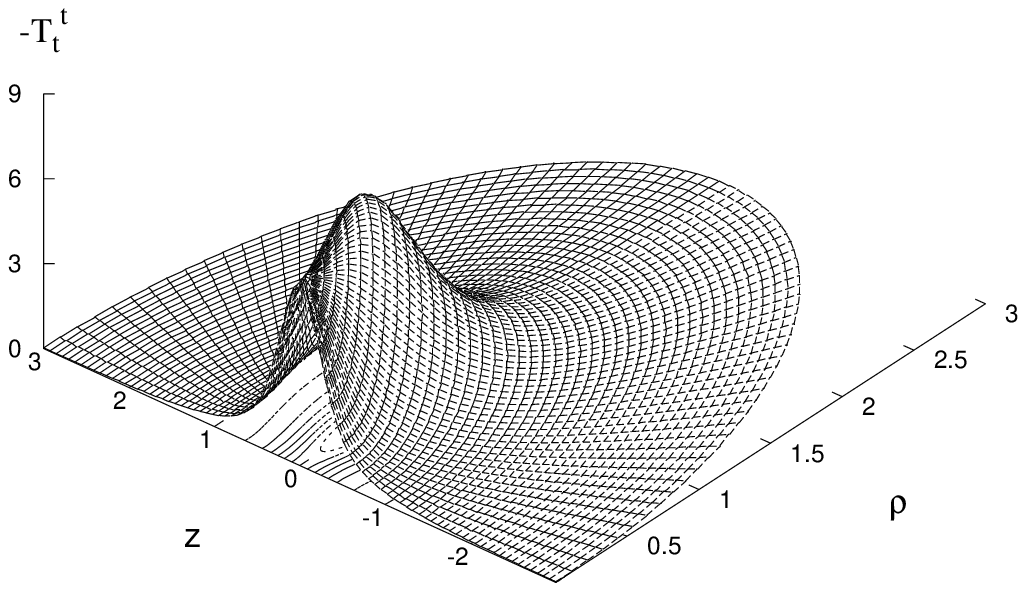,width=12cm}}
\end{picture}
\\
\\
{\small {\bf Figure 3.}
The dimensionless energy density of a $n=2$ MM solution in a fixed AdS
background with $\Lambda=-1$ is plotted as a function of $\rho$ and $z$.
The solution has been found for a Higgs coupling constant $\lambda=10$
and has a  mass $M=3.998$.}
\\
\\
Qualitatively, the behaviours of the Higgs field and Yang-Mills fields
are very similar to that corresponding to Minkowski spacetime monopoles.
In particular, we notice a similar shape for the functions
$H_i$ and $\Phi_i$ and also for the energy density.
Again, the functions $H_2,~H_4$ and $\Phi_1$ present a small
$\theta-$dependence. For this type of solutions, the nodes of the Higgs
field are superposed at the origin.

In Figure 1 the mass  of the solutions in units of $4\pi \eta/g$
is plotted as a function of the cosmological constant
for various winding numbers and a fixed value of the coupling constant
$\lambda=0.5$.

The energy of the multimonopole is of the order of
$n$ times the corresponding one-monopole energy.
However, for any value of $\Lambda$, the ratio $M/n$ increasing with
increasing $n$. Also, this quantity is greater than
the mass of
the corresponding $n=1$ spherically symmetric monopoles.
For example, for solutions with $\Lambda=-4$ and $\lambda=0.5$, we find
$M(n=2)/M(n=1)=2.466,~~M(n=3)/M(n=1)=4.373$ while $M(n=4)/M(n=1)=6.678$.

Most of our results have been obtained in the Prasad-Sommerfeld limit,
but similar conclusions hold for a nonvanishing Higgs self-coupling
constant $\lambda$. To demonstrate this dependence,
we plot in Figure 2 the mass of
solutions per topological charge
as a function of $\lambda$ for two different values of the cosmological
constant and winding numbers $n=1,~2$.

For all configurations, the energy density $\epsilon=-T_t^t$ of the
solutions has a strong peak along the $\rho$ axis,
and it decreases monotonically along the symmetry axis. Individual
unit charged monopole components of the MM do not feature distinct peaks
(see Figure 3).
Equal density contours reveal a torus-like shape of the solutions.
For increasing $|\Lambda|$, these peaks increase in size
and the energy density becomes localised in a decreasing region of space.
\subsection{Monopole-antimonopole solutions}

As discussed in \cite{Taubes:1982ie} for $n=1$ and $\Lambda=0$, there
exist also a different type of solutions of the second order
Euler-Lagrange equations, which are not stable and represent
saddle points of the energy, rather than absolute minima resulting from
solving the Bogomol'nyi equations. Here we construct the simplest
examples of such 
solutions, namely MA solutions obtained by taking
$m=2$ in the general Higgs field asymptotics at infinity,
\re{evenm}-\re{BC-H},
which implies the asymptotic behaviour 
$H_1=H_3=0,~~H_2=H_4=-1.$
This corresponds to a MA configuration, with a vanishing
net magnetic charge.

Similarly to the multimonopole case, we need to consider solutions
only in the region $0 \leq \theta \leq \pi/2$;
on the $z$- and $\rho$-axis the functions
$H_1, H_3,  \Phi_2$ and the derivatives
$\partial_\theta H_2,\partial_\theta H_4$ and $\partial_\theta \Phi_1$
are to vanish. The same gauge condition, \re{gauge},
fixes the residual Abelian gauge invariance.

A calculation similar to that done in \cite{Kleihaus:2000sx}
shows that this configuration
possesses two magnetic charges of opposite sign, located
on the positive and negative $z-$axis, respectively, with
nonvanishing local density of magnetic charge.
However, the integral (\ref{topch}), evaluated at infinity,
for a surface enclosing both charges, yelds a zero net magnetic charge.
Also, the expansions of the matter functions
at the origin and on the $z-$axis presented in Ref.
\cite{Kleihaus:2000sx} are still valid, and will not be given here.

The magnetic dipole moment $C_m$ of these solutions can be obtained from
the asymptotic form of the non-abelian gauge field, after
choosing a gauge where the Higgs field is
asymptotically constant $\Phi \to \tau_3$,
which yields
\begin{eqnarray}
\label{mon}
A_{k}dx^{k}= C_m\frac{\sin^2 \theta}{r}\frac{\tau_3}{2} d\varphi.
\end{eqnarray}

We have constructed MA solutions for a large range of the parameters
$(\Lambda,~\lambda$). For vanishing $\Lambda$, our results are in very
good agreement with those of \cite{Kleihaus:2000sx}.
The main part of the numerical analysis has been done for the
case $n=1$, the case $n=2$ being studied more briefly.
However, most of the qualitative properties of the $n=2$ solutions do not
differ from those of the
$n=1$ case.

Similarly to the case of multimonopoles, we have found that a  negative
cosmological constant does not change the qualitative properties of the
solutions. As expected, a negative cosmological constant 
affects the behaviour of the fields in the asymptotic region,
where the modulus of the Higgs field, for example, reaches its
asymptotic value faster.

In Figure 4 we present the
the mass and the magnetic dipole moment of $n=1$ MA solutions
as a function of $\Lambda$ for three values of the
Higgs self-coupling constant $\lambda$.

The  energy density  always possesses maxima on the positive and negative
$z$-axis at the locations of the monopole and antimonopole and a saddle
point at the origin. An increase of $|\Lambda|$ makes the maxima of the
energy density higher and sharper.
At the same time, the modulus of the Higgs field tends
faster towards its vacuum expectation value.
The typical energy density for nongravitating MA solutions
have a similar form to
that presented in Section 4 for gravitating configurations.
\newpage
\setlength{\unitlength}{1cm}
\begin{picture}(18,7.5)
\centering
\put(2,0){\epsfig{file=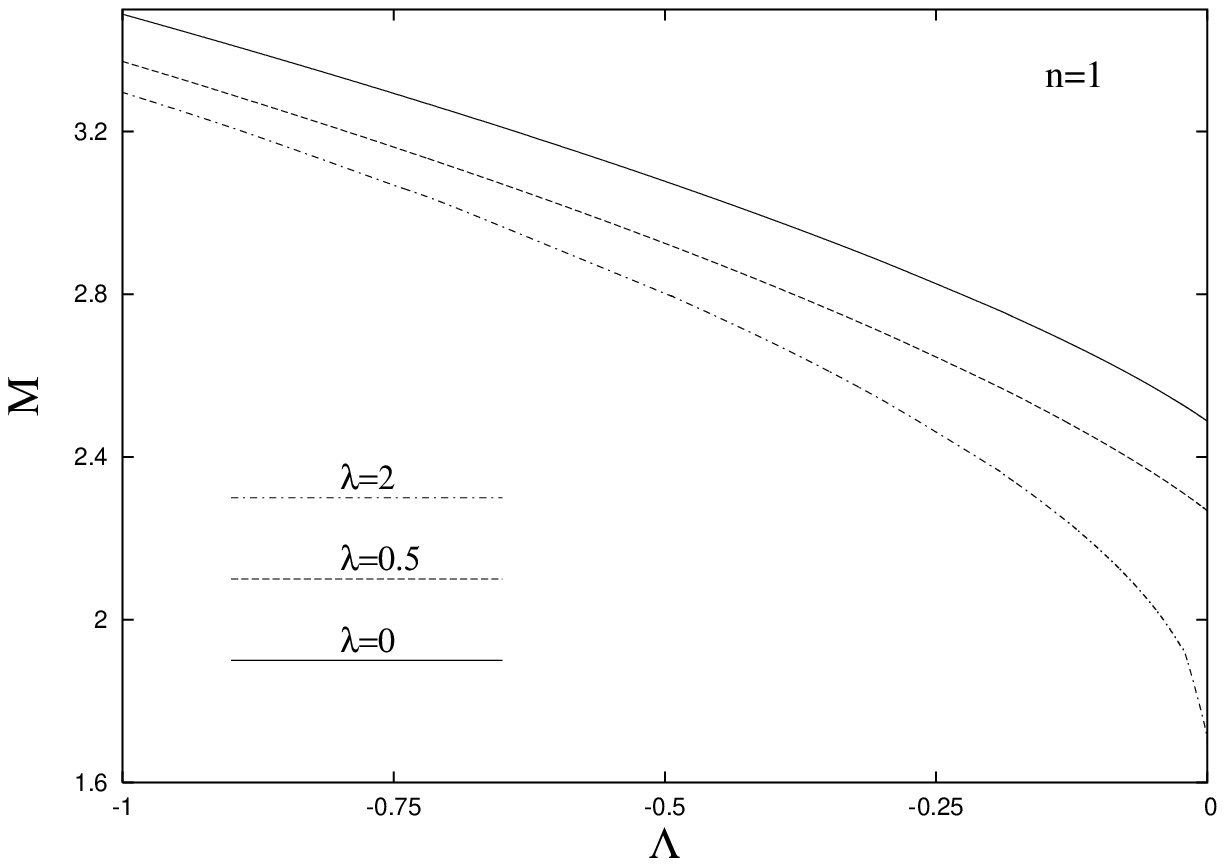,width=11.cm}}
\end{picture}
\\
\\
\setlength{\unitlength}{1cm}
\begin{picture}(18,8)
\centering
\put(2.5,0){\epsfig{file=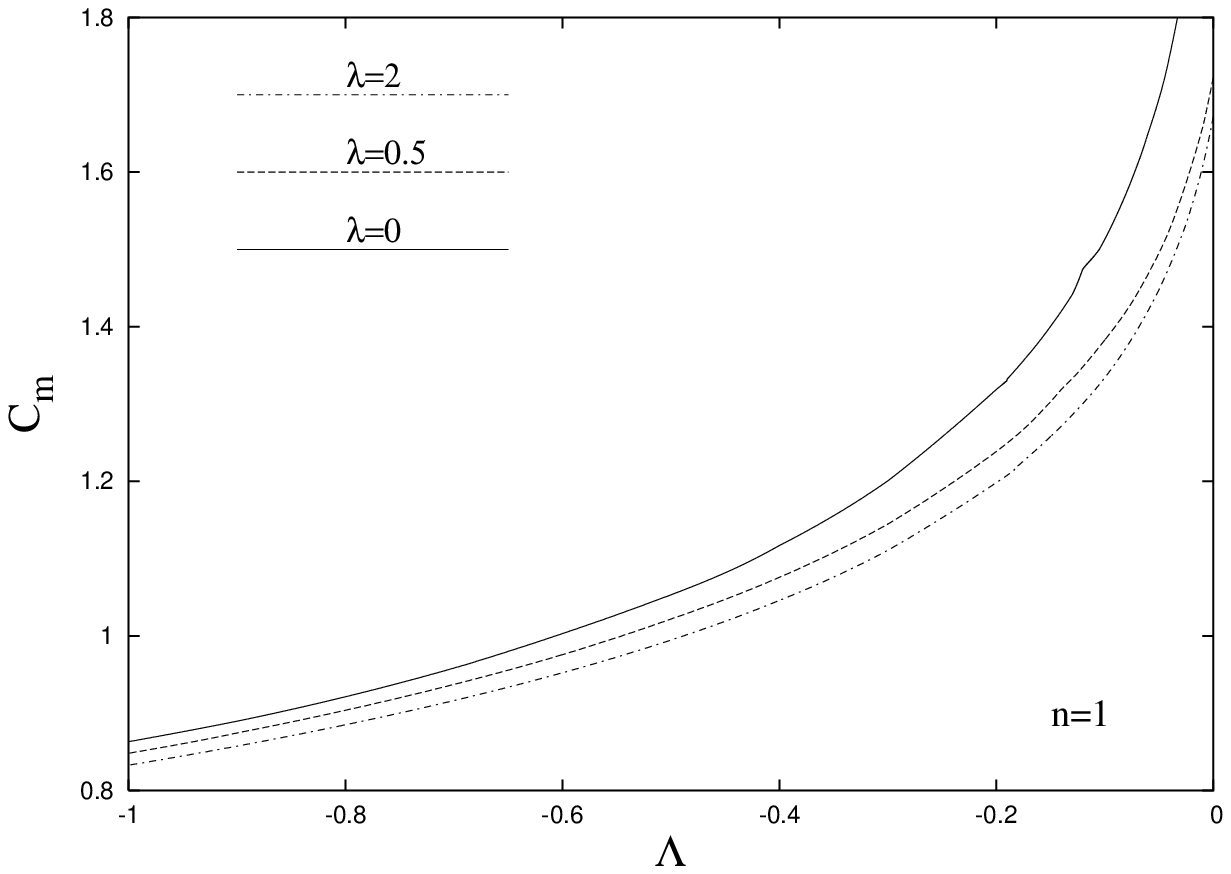,width=11.cm}}
\end{picture}
\\
\\
{\small {\bf Figure 4.}
The mass and the magnetic dipole moment of $n=1$ MA solutions is plotted
as a function of $\Lambda$ for three values of the
Higgs self-coupling constant $\lambda$.}
\\
\\
The nodes  of the modulus of the Higgs field  correspond to the locations
of the particles. With increasing $|\Lambda|$, the distance $d$ between
the MA centers becomes smaller tending to a finite limit as
$\Lambda \to -\infty$, making it more difficult to distinguish the
monopole from antimonopole.
Several values we found
for solutions with $\lambda=2$ are:
$d(\Lambda=0)=3.24,~d(\Lambda=-0.5)=1.52,~d(\Lambda=-1)=1.22,
~d(\Lambda=-2.5)=0.9$, the distance between
the MA centers approaching
for larger values of $|\Lambda|$ a limiting value $d\simeq 0.85$.
The magnetic dipole moment $C_m$ decreases in the same limit.

Considering solutions with a fixed negative value
of $\Lambda$, we find that the distance between
the MA centers decreases as $\la$ increases and converges to a
finite, nonzero value.
For example, for MA solutions 
in an AdS spacetime with $\Lambda=-1$, $d$ decreases from
$d(\lambda=0)=0.68$ to a limiting value $d\simeq 0.6$. This property is
shared with the $\Lambda=0$ case \cite{Kleihaus:2000sx}.


\section{Inclusion of gravity}
We now consider the effects of these axially symmetric configurations
on the AdS spacetime, by including 
the Einstein gravity term with a negative cosmological constant
$\sqrt{-g}(-R/(16 \pi G)+2\Lambda)$ in the action density in (\ref{lag0}).

The usual rescaling (\ref{resc}) reveals in this case the existence
of a new dimensionless coupling constant
$\alpha^2=4 \pi G \eta^2$, where $G$ is the Newton constant.
The complete classification of the solutions in the
space of physical parameters $(\alpha,~\Lambda,~\lambda)$
is a considerable task which is beyond the scope of this paper.
Instead, we analysed the situation for several values of $\Lambda$ which,
hopefully, reflect all the properties of the general pattern.
Also, for simplicity we will study in this section
only solutions with no Higgs potential  ($\lambda=0$). 

The asymptotically anti-de Sitter (AAdS) 
gravitating counterparts of the  axially symmetric configurations
discussed in Section 3 are found by using a metric of the form
\begin{eqnarray}
\label{ansatz-axial}
ds^2= \frac{m}{f} ( \frac{d r^2}{1-\frac{\Lambda}{3}r^2}+ r^2 d \theta^2 )
+ \frac{l}{f} r^2 \sin ^2 \theta d\phi^2 - f(1-\frac{\Lambda}{3}r^2) dt^2,
\end{eqnarray}
with $f$, $l$  and $m$ being functions of $r$ and $\theta$.
To obtain  asymptotically AdS regular solutions with finite energy,
the metric functions have to satisfy the boundary conditions
$f= m= l =1$ at infinity, and $\partial_r f=\partial_r m=\partial_r l= 0$
at the origin. The boundary conditions on the $z$-axis are
$\partial_\theta f= \partial_\theta m=\partial_\theta l=0$, which are
consistent with the requirement of invariance under reflection in the
$\theta=\pi/2$ plane. Finally, regularity on the $z$-axis requires also
$m=l$ for $\theta=0$. These metric boundary conditions are valid for both
MA and MM gravitating configurations.
The boundary conditions for the gauge potentials and
Higgs field are similar to the nongravitating case.

To solve the set of nine EYMH equations we employ the same numerical
algorithm as for the YMH solutions in fixed AdS background. In the
numerical procedure we use a suitable combination of the Einstein
equations
\begin{equation}
\label{einstein-eqs}
R_{\mu\nu}-\frac{1}{2}g_{\mu\nu}R +\Lambda g_{\mu\nu} = 8\pi G T_{\mu\nu},
\end{equation}
such that the diferential equations for metric variables $(f,l,m)$ are
diagonal in the second derivatives with respect to $r$.

Similar to the asymptotically flat case, the value of the mass is encoded
in the asymptotics of the metric functions $f,l,m$,
whose expression, valid for large $r$, is
\begin{eqnarray}
\label{asimpt-f}
f=1+\frac{f_1+f_2 \sin^2 \theta}{r^3}+O(\frac{1}{r^5}),~~
m=1+\frac{m_1+m_2 \sin^2 \theta}{r^3}+O(\frac{1}{r^5}),~~
l=1+\frac{l_1+l_2 \sin^2 \theta}{r^3}+O(\frac{1}{r^5}),
\end{eqnarray}
where $f_1,~f_2$ are constants to be determined numerically, and
\begin{eqnarray}
\label{test}
l_1=m_1=\frac{2f_1}{3},~~~l_2=\frac{6f_2}{17},~~~m_2=\frac{14f_2}{17}.
\end{eqnarray}
The generalization of Komar's formula for  AdS asymptotics is not
straightforward and requires the further subtraction of a background
configuration in order to yield a finite result.

To compute the mass of these configurations, we  employ
the counterterm formalism proposed by
Balasubramanian and Kraus \cite{Balasubramanian:1999re}.
This technique was inspired by AdS/CFT correspondence and consists of
adding suitable counterterms $I_{ct}$ to the total action of the theory.
These counterterms are built of curvature invariants of a boundary metric
and thus obviously do not alter the bulk equations of motion.
However, they do regularise the boundary stress tensor and also the
conserved charges. The details of the computation, which we omit here, are
presented in \cite{Radu:2004gu}, where an axially symmetric line element
with the same parametrisation and the same asymptotics is considered.

The expression of the mass derived in this way is
\begin{eqnarray}
\label{ct-mass}
M= \frac{\Lambda}{3G}\left(\frac{2f_1}{3}+\frac{8f_2}{17}\right).
\end{eqnarray}
The same form is obtained by using the Hamiltonian formalism of Henneaux
and Teitelboim \cite{Henneaux:1985tv}.

\newpage
\setlength{\unitlength}{1cm}
\begin{picture}(18,7.5)
\centering
\put(2,0){\epsfig{file=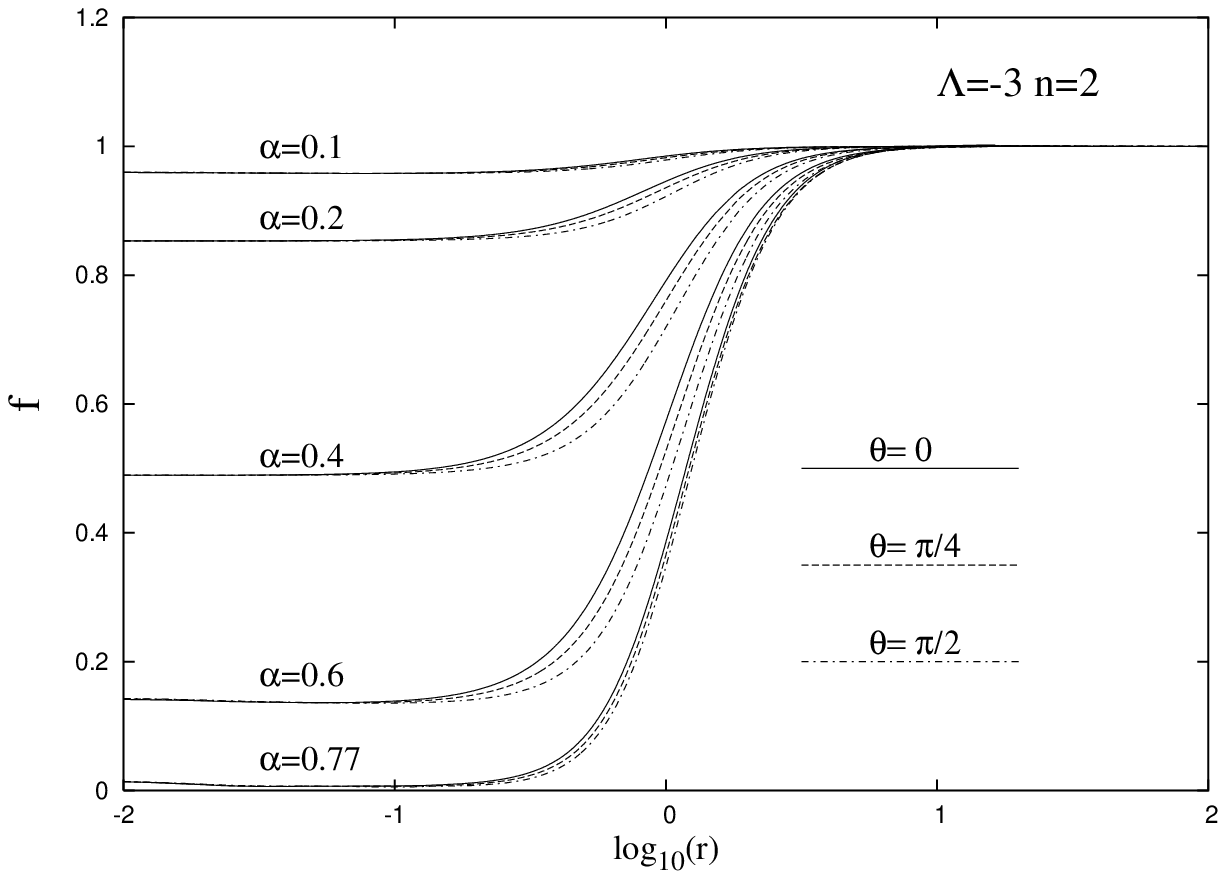,width=12cm}}
\end{picture}
\\
\\
\setlength{\unitlength}{1cm}
\begin{picture}(18,8)
\centering
\put(2.5,0){\epsfig{file=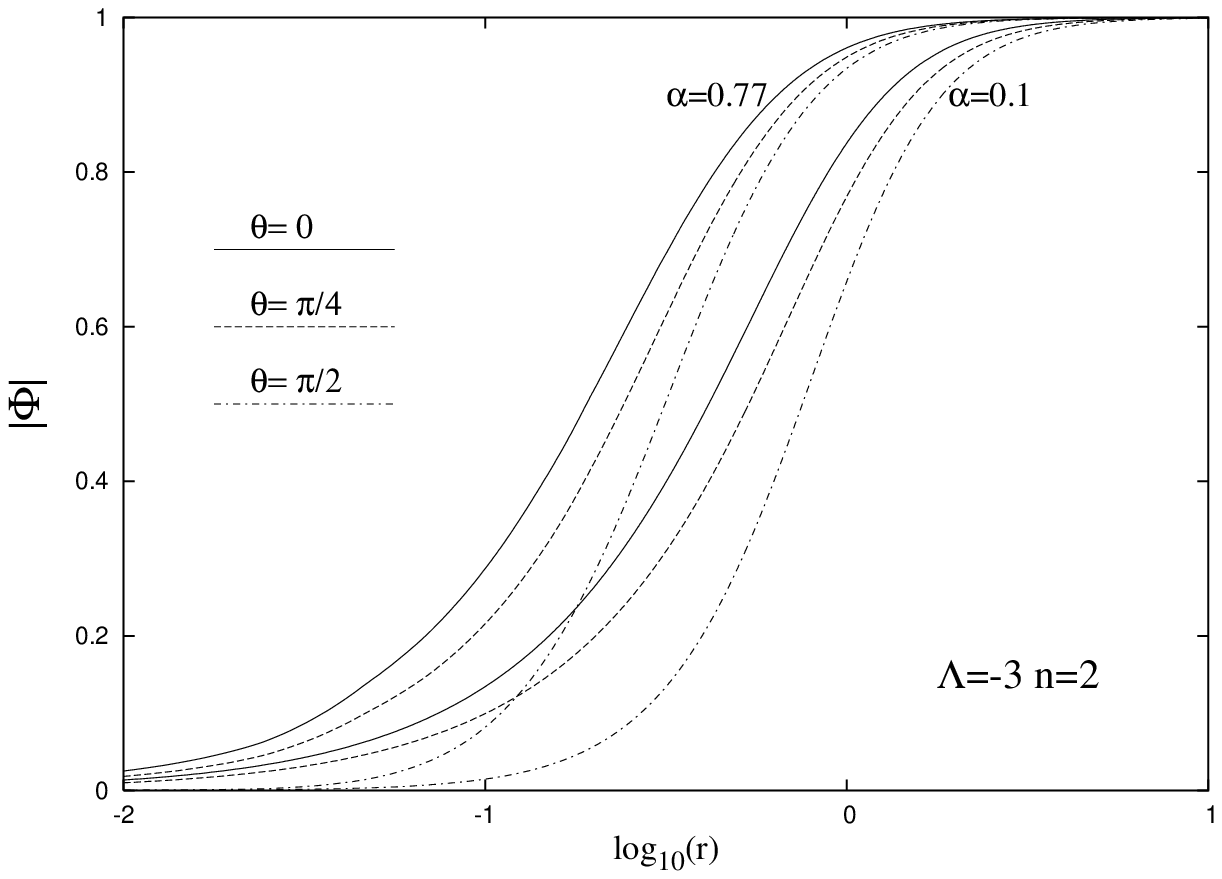,width=12cm}}
\end{picture}
\\
\\
{\small {\bf Figure 5.}
The metric function $f$ and the norm of the Higgs field
$|\Phi|=\sqrt{\Phi_1^2+\Phi_2^2}$ of typical AAdS MM solutions
are shown as a function of $r$
for three values of the angle $\theta$  with $n=2$, $\Lambda=-3$
and a sequence of values of $\alpha$.}
\\
\\
The dimensionless mass is given by
\begin{eqnarray}
\label{adim-mass}
\mu/\alpha^2=\frac{e}{4 \pi \eta}M.
\end{eqnarray}
\subsection{Gravitating MM}
A spherically symmetric line element is obtained for $l=m$, with $f$ and
$m$ being functions of the coordinate $r$ only. Some properties of the
corresponding $n=1$ monopole solutions are discussed in \cite{Lugo:1999ai}
by using Schwarzschild-like coordinates. Similar results are found also
by using the above metric parametrisation. 

As found in \cite{Lugo:1999ai},
when $\alpha$ is increased from zero, while $\Lambda $ is kept fixed, a
gravitating monopole branch emerges
\newpage
\setlength{\unitlength}{1cm}
\begin{picture}(18,7.5)
\centering
\put(2,0){\epsfig{file=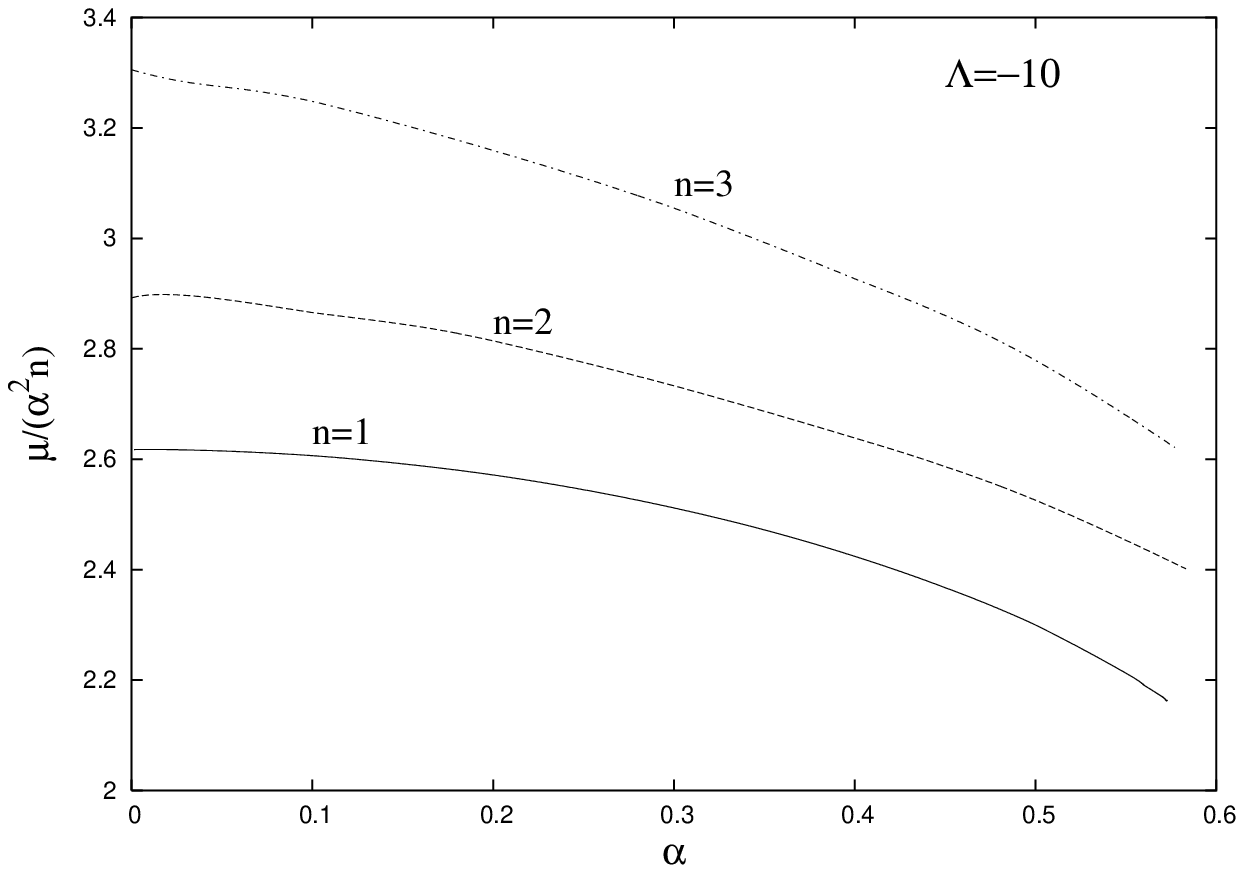,width=12cm}}
\end{picture}
\\
\\
{\small {\bf Figure 6.}
The mass per topological charge is plotted as a function of
$\alpha$ for $n=1,~2,~3$ (multi-)monopole solutions with  $\Lambda=-10$
and $\lambda=0$.}
\\
\\
smoothly from the corresponding AdS
space monopole solutions. The total mass of  gravitating solutions
decreases with  increasing $\alpha$. A similar property has been noticed
\cite{Breitenlohner:} for asymptotically flat solutions too.

The fundamental $n=1$ monopole branch extends up to a maximal
value $\alpha_{max}$ of the coupling constant $\alpha$.
This maximal value decreases with $|\Lambda|$; for example
$\alpha_{max}(\Lambda=0)\simeq 1.4$,
$\alpha_{max}(\Lambda=-0.1)\simeq 1.28$, while
$\alpha_{max}(\Lambda=-3)\simeq 0.80$.
As is known from the work of \cite{Breitenlohner:},
for $\Lambda=0$, and with the Higgs coupling $\lambda=0$, the fundamental
monopole branch bends backwards at $\alpha_{max}$, until a critical value
$\alpha_{cr}$ corresponding to the extremal 
Reissner-Nordstr\"om (RN) solution
is reached. Fro $\Lambda<0$, we refer to such RN solutions as RNAdS.
However, we have
found that for large enough values of
$|\Lambda|$, the maximal value $\alpha_{max}$ and the critical
value  $\alpha_{cr}$ seem to coincide, as they do also for the
asymptotically flat case when $\lambda >0$~\cite{Breitenlohner:}.

Along the fundamental branch, the metric function $f(r)$ develops
a minimum, which decreases asymptotically to zero, as
$\alpha \to \alpha_{cr}$. The functions $\omega(r)$ and $H(r)$
parametrising the gauge and the Higgs fields respectively, approache
their respective RNAdS values $\omega=0,~H=1$~\cite{Lugo:1999ai}.
It is also noticed that the critical value of $\alpha$ as well as the
corresponding value of mass decrease with $\Lambda$.
The limiting spacetime consists of an inner part with $r<r_c$ and an
outer part with $r\ge r_c$. The exterior of the critial solution
corresponds to the exterior of a RNAdS black hole solution with a
degenerate horizon at $r=r_c$ and unit magnetic charge.
A  discussion of these solutions and typical profiles are
presented in \cite{Lugo:1999fm,Lugo:1999ai}.

Here we have constructed axially symmetric multimonopoles for
$n>1$, the metric functionsin this case displaying an angular dependence.
Most of the properties of these solutions are similar to those
corresponding to asymptotically flat space. In particular,
for a given topological charge $n$, we find a branch
of globally regular asymptotically AdS multimonopoles, emerging
smoothly from the corresponding solutions in fixed AdS background.

In Figure 5 we show the metric function $f$ and the norm of the Higgs
field $|\Phi|=\sqrt{\Phi_1^2+\Phi_2^2}$ of gravitating $n=2$ multimonopole
solutions with $\Lambda=-3$, for several values of $\alpha$.

The fundamental branch extends up to some maximal value of $\alpha$,
whose precise value depends on $\Lambda$. Along this fundamental branch,
the mass of the solutions decreases monotonically.
For the studied case $\lambda=0$, the value of $\alpha_{max}$
decreases with increasing $\Lambda$. The value at the origin of the
metric function $f$ decreases with increasing $\alpha$, corresponding as
$\alpha \to \alpha_{cr}$ to the degenerate horizon of a RNAdS solution.
At the same time the Higgs field function $\phi_1$ approaches the unit
value. The exterior of the critical solution corresponds to the exterior
of a degenerate horizon RNAdS black hole with magnetic charge $n$
(note that for the same value of
$\alpha$ and $n_2>n_1$, the mass of this RNAdS black hole
satisfies the relation $M(n_2)>M(n_1)$).

\newpage
\setlength{\unitlength}{1cm}
\begin{picture}(18,6)
\centering
\put(2,0){\epsfig{file=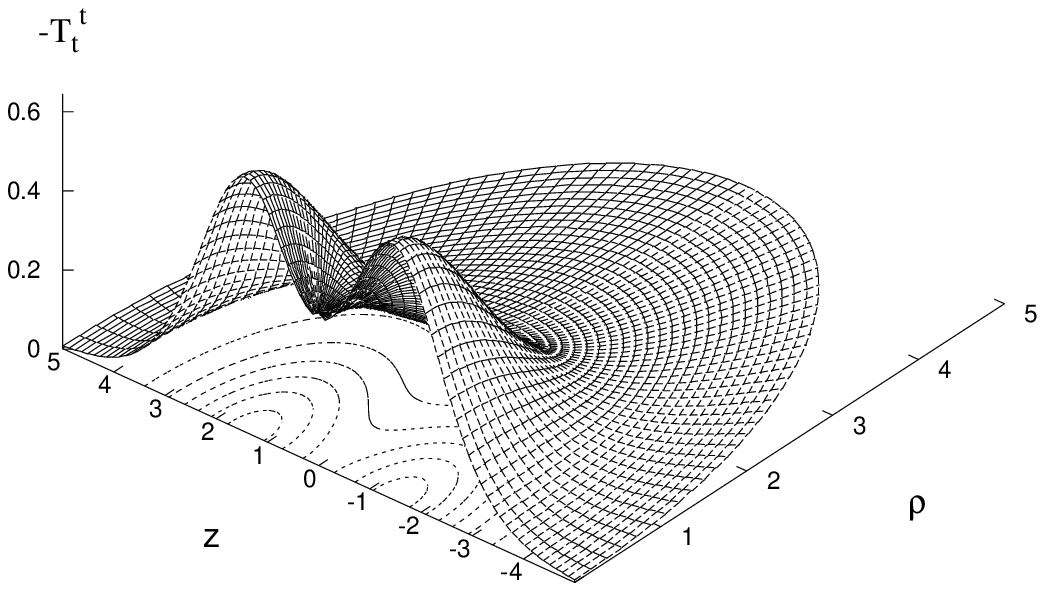,width=12cm}}
\end{picture}
\\
\\
{\small {\bf Figure 7.}
The dimensionless energy density of a $n=1$ gravitating MA solution
with $\Lambda=-0.04$, $\alpha=0.2$,
is plotted as a function of $\rho$ and $z$.
}
\\
\\
The mass per unit charge of the multimonopole solutions decreases with
increasing $\alpha$. As discussed in \cite{gam}, the mass {\it per unit
charge} $n$ of the $\Lambda=0$ multimonopoles is smaller than the mass of
$n=1$ monopoles, {\it i.e.} that like monopoles attract.
We have found however that this is not a generic property of AAdS
solutions, and that for large enough values of $\Lambda$ the
monopoles of like charge in a gravitating Higgs model on AdS space
repel for all values of $\alpha$ (a typical situation is presented
in Fig. 6).
\subsection{Gravitating MA}
For completeness, we present here some properties of the gravitating
MA solution with $\Lambda<0$. The field equations have been solved in
this case by using the same metric ansatz (\ref{ansatz-axial}), and taking
again $m=2$ in the matter field asymptotics at infinity,
\re{evenm}-\re{BC-H}, for $n=1$,~$\lambda=0$ and several values of
$\Lambda$.

For MA boundary conditions, we find that, similar to the MM case, a branch
of gravitating MA solutions emerges smoothly from the MA solutions in
fixed AdS background. The basic properties of these solutions are similar
to those of the configurations in a fixed AdS background.
For example, the modulus of the Higgs field always possesses two zeros on
the $z-$axis, corresponding to the location of the monopole and the
antimonopole.

With increasing $\alpha$,
 the monopole and antimonopole
move closer to the origin and the mass $\mu$
of the solutions decreases.
In Figure 7 we plot the  energy density $\epsilon=-T_t^t$
of a typical MA solution as a function of the
coordinates $\rho, z$, for $\Lambda=-0.04$, $\alpha=0.2$.

This fundamental branch of solutions ends at a critical value
$\alpha_{cr}$, when gravity becomes too strong for solutions to persist.
As expected, this value decreases with $|\Lambda|$  (for example
$\alpha_{cr}(\Lambda=0)\simeq0.67$ \cite{Kleihaus:2000hx},
$\alpha_{cr}(\Lambda=-0.01)\simeq 0.64$ while for $\Lambda=-1$ we find
$\alpha_{cr}\simeq 0.35$).

In asymptotically flat space, it has been found that at $\alpha_{cr}$
a second branch of MA solutions emerges, extending back to $\alpha=0$
\cite{Kleihaus:2000hx}. Along this upper branch, the distance $d$ between
the MA centers shrinks to zero size in the limit $\alpha \to 0$. After a
suitable rescaling, as $\alpha \to 0$ the configuration
approaches the spherically symmetric Bartnik-McKinnon (BK) solution, with
$H_1 =H_3 =0,~~H_2 =H_4 =\omega(r)$, with $\omega(r)\to -1$ asymptotically
(thus $\alpha \to 0$ means here $\eta \to 0$ with fixed $G$).
In AdS spacetime however, the situation is to be more complicated since
the BK configuration is now replaced by a continuum of solutions.
Here the picture depends crucially on the value of $\Lambda$ (in
particular, solutions with $\omega\to -1$ cease to exist for a large
enough $|\Lambda|)$, and the meaning of the limiting solutions is less
clear. We hope to come back to this point in future.


\section{Conclusions}

We have studied, analytically and numerically, the basic properties of
two types of solutions to the $SU(2)$ Yang-Mills-Higgs model 
in the presence of a negative cosmological constant.
The first type of solutions studied are the axially symmetric
multimonopoles (MM) with magnetic charge equal to the azimuthal winding
number $n$. The second type are the axially symmetric
monopole--antimonopole (MA) pairs with vanishing magnetic charge and
azimuthal widing number $n$, and nonvanishing magnetic dipole moment.
These are the simplest examples of MA chains and vortex lines, and are
not topologically stable even in flat spacetime, unlike the MM solutions.
Both MM and MA solutions obey the same Euler--Lagrange equations of this
system, subjected to axial symmetry. The drastic difference in the said
properties arises from the imposition of different boundary conditions at
infinity, in each case, and while the MM solutions are labeled by an
{\it azimuthal winding number} $n$, the MA solutions are further labeled
by a {\it number} $m$ multiplying the boundary value of the {\it polar}
angle, {\it i.e.} by the pair of integers $(n,m)$. We have emphasised this
point and have presented a unified treatment of both types of solutions.

The spherically symmetric analogues of the MM solutions studied here
were discussed in \cite{Lugo:1999fm, Lugo:1999ai}, where unit magnetically
charged monopoles on both fixed AdS backgrounds and gravitating ones were
considered, but not MA type solutions which are at most axially symmetric.

Qualitatively, the behaviour of the AdS solutions we constructed were
found to be very similar to that of corresponding
Minkowski spacetime configurations. Thus, it seems that when studying
gauge field systems containing a scalar field, the Higgs field in the case
at hand, the solutions exhibit a generic behaviour for $\Lambda \leq 0$.
A similar behaviour has been noticed for sphalerons
\cite{VanderBij:2001ah}.

Let us list some points of contrast and similarity of MM and MA
solutions, between the $\Lambda=0$ and $\Lambda<0$ cases:
\begin{itemize}
\item
With $|\Lambda|>0$, the MM solutions satisfy the second order
Euler--Lagrange equations and not the first order Bogomol'nyi equations,
even when the Higgs potential is absent, i.e. when $\la=0$ in \re{lag0}.
The immediate consequence of this is that in this limit ($\la=0$), like
charged monopoles are not non--interacting. It turns out that the mass
of one $n=2$ MM is larger than that of two $n=1$ MM's with their
centres infinitely far apart. Thus, it looks as if the $n=2$
axially symmetric MM is unstable since like monopoles repel each other.
This feature holds for all $n$, and is in contrast to the MM
solutions in flat spacetime, which are stable for any $n$ since they
are all self--dual solutions.
\item
When $|\Lambda|>0$, it turns out that like monopoles remain mutually
repulsive even for positive Higgs coupling constant $\la>0$ model.
This property, which is not surprising, is similar to the flat
spacetime, $\Lambda=0$ case. Like in the case of flat background
moreover, when $\la\gg 1$, the mass of the MM solution tends to a finite
limiting value~\cite{Kirkman:1981ck} which we have not estimated here.
This may be an interesting detail to return to in future.
\item
It is known from \cite{Kleihaus:2000sx} that in the flat space model the
distance on the $z$-axis between the monopole and the antimonopoled
ecreases as the Higgs coupling constant $\la$ increases, converging to
a finite limit. In the present model with $|\Lambda|\neq 0$, this
property is similar for the MA solutions. For a fixed negative value
of the cosmological
constant, this distance decreases as $\la$ increases and converges to a
finite value. This is qualitatively the same as for the flat space model,
but quantitatively this distance decreases faster with increasing $\la$.
\item
For a fixed value of the Higgs coupling constant $\la$, increasing
$|\Lambda|\neq 0$ of the negative cosmological constant results in the
decrease of the distance between the monopole and the antimonopole of
the MA solution, converging to a finite value.
\item
Another one of the properties of in Minkowski space that persists in the
AdS case, is that found by Houston and O'Raifeartaigh~\cite{Houston:aj}:
Any regular axially symmetric magnetic charge distribution
can be located only at isolated points situated on the axis of symmetry,
with equal and opposite values of the charge at alternate points.
In particular, if only one sign of the charge is allowed,
all the charge must be concentrated at a single point. The results of the
present paper support an extension of this property to the AdS case.
\end{itemize}

As expected, the inclusion of gravity  does not change the general
picture, since for $\Lambda<0$, the gravitational field does
not affect the behaviour of the YMH system at infinity.
As it happens in asymptotically flat case,
a critical value for the Newton constant exists above which no regular
solution can be found. However, as in the spherically symmetric case,
this critical value is   smaller than the corresponding
value of the asymptotically flat one.

For asymptotically flat solutions, the inclusion of gravity allows
for an attractive phase of like monopoles not present in flat space.
Here we have presented numerical arguments that for large enough values
of $|\Lambda|$, only a repulsive phase exists for like monopoles.

Hairy black hole solutions of EYMH theory with a negative cosmological
constant, generalising  the asymptotically flat configurations discussed
in \cite{gam, Hartmann:2001ic} can also be constructed numerically, but
this was not done in the present work..

One may ask about the possible relevance of
these solutions within the holographic principle and its AdS/CFT
correspondence realisation.
A scalar field has been discussed in this context by many authors.
For example, the asymptotic behaviour of the bulk
scalar field was directly related to
one-point functions in the dual CFT \cite{Balasubramanian:1998de}.
The vortex solution of the Abelian Higgs model has also an interesting
AdS/CFT interpretation, its mass density being dual to the discontinuity
of the logarithmic derivative of the correlation function of the boundary
scalar operator \cite{Dehghani:2001ft}. To date however the crucial
interplay between the non Abelian gauge fields and a scalar multiplet
(with its rich set of boundary conditions discussed in this paper)
is not considered in this context in the literature.

On the other hand, the EYMH bulk action (or a suitable modification
supporting the same set of asymptotic boundary conditions) does
not seem to correspond to a known supergravity truncation,
and in particular we do not know the underlying boundary CFT.
We believe that further work in this direction will be of interest.
\\
\\
{\bf Acknowledgement}
\\
This work was carried out in the framework of Enterprise--Ireland
Basic Science Research Project SC/2003/390.


\end{document}